\documentclass[conference]{IEEEtran}
\usepackage{amsmath, amssymb, bm, cite, epsfig, psfrag}
\usepackage{graphicx}
\usepackage{soul} 
\usepackage[margin=0.530in]{geometry}
\usepackage{dblfloatfix}
\usepackage{array}
\usepackage{lipsum}

\usepackage[font=small]{caption}
\usepackage{epstopdf}	
\usepackage{longtable}
\usepackage{supertabular,booktabs}
\usepackage{bbm}
\usepackage{multirow}
\usepackage[usenames,dvipsnames]{xcolor}

\usepackage{subfigure}
\usepackage{etoolbox}
\usepackage{pbox}
\usepackage{fixltx2e}
\usepackage{tabu}	
\usepackage{filecontents}
\usepackage{enumerate}
\usepackage{textcomp}
\usepackage{colortbl}
\usepackage{fancyhdr}
\usepackage{bm}

\newtoggle{conference}

\togglefalse{conference} 
\graphicspath{{figures/}}

\newcolumntype{?}{!{\vrule width 2pt}}

\newcolumntype{P}[1]{>{\centering\hspace{0pt}}p{#1}}
\newcolumntype{M}[1]{>{\centering\hspace{0pt}}m{#1}}
\newcolumntype{L}{>{\centering\arraybackslash}m{3cm}}

\usepackage[utf8]{inputenc}

\fancypagestyle{firststyle}{ 
	\fancyhf{}
	\fancyhead[L]{\vspace{+0.2 cm} Y. Xing and T. S. Rappaport, ``Terahertz Wireless Communications: Co-sharing for Terrestrial and Satellite Systems above 100 GHz (Invited Paper),'' submitted to 2021 \textit{IEEE Communications Letters}, May 2021, pp. 1-5.}

}
\IEEEoverridecommandlockouts
\begin{document}
\title{Terahertz Wireless Communications: Co-sharing for Terrestrial and Satellite Systems above 100 GHz}
 \vspace{-1.40 cm}
\author{\IEEEauthorblockN{Yunchou Xing and Theodore S. Rappaport (Invited Paper)}
\IEEEauthorblockA{\small NYU WIRELESS, NYU Tandon School of Engineering, Brooklyn, NY, 11201,
				\{ychou, tsr\}@nyu.edu}\vspace{-0.7cm}
					\thanks{This research is supported by the NYU WIRELESS Industrial Affiliates Program and NSF Research Grants: 1909206 and 2037845.}
}
\maketitle
\thispagestyle{firststyle}
\begin{abstract} 
This paper demonstrates how spectrum up to 1 THz will support mobile communications beyond 5G in the coming decades. Results of rooftop surrogate satellite/tower base station measurements at 140 GHz show the natural isolation between terrestrial networks and surrogate satellite systems, as well as between terrestrial mobile users and co-channel fixed backhaul links. These first-of-their-kind measurements and accompanying analysis show that by keeping the energy radiated by terrestrial emitters on the horizon (e.g., elevation angles $\leq$15\textdegree), there will not likely be interference in the same or adjacent bands between passive satellite sensors and terrestrial terminals, or between mobile links and terrestrial backhaul links at frequencies above 100 GHz. 
\end{abstract}

\begin{IEEEkeywords}                            
mmWave; Terahertz; spectrum sharing and coexistence; satellite;  OOBE; interference mitigation;  \end{IEEEkeywords}


\section{Introduction}~\label{sec:intro}
As the wireless world moves towards future generations of communications (6G and beyond), new applications such as wireless cognition (providing human intelligence over wireless communications \cite{rappaport19access,ghosh195g}), will require data rates on the order of hundreds of Gbps or even Tbps with near-zero latency. To accommodate such massive data rates, 6G and beyond will likely flourish at frequencies above 100 GHz (e.g., sub-THz bands of 100-300 GHz or THz bands of 300 GHz-3 THz) as electronics become available to exploit the abundant spectrum \cite{rappaport19access,viswanathan20A,marcus21a,Hadeel20THz}. This new era, which will exploit wide-bandwidth channels (e.g., 1 GHz or more) and adaptive antenna arrays in small form factors, will enable the proliferation of new applications such as centimeter-level position location, high-resolution virtual/augmented reality (VR/AR), unmanned aircraft system (UAS) and high altitude platform stations (HAPS), driver-less cars, factory automation, remote medicine, radars for transportation and motion sensing, spectroscopy, sensing, imaging, and fixed and mobile broadband wireless \cite{rappaport19access,viswanathan20A, ghosh195g, Hadeel20THz}.


Global regulatory bodies and standard agencies govern the use of radio frequencies (mainly below 275 GHz) to promote efficient spectrum use. Specific provisions on frequencies above 100 GHz \cite{marcus19a,marcus21a} were instituted by the Japanese regulator, Ministry of Internal Affairs and Communications, in 2015 at 116-134 GHz \cite{JapanmmWave}, and the group of European spectrum regulators, Conference of European Postal and Telecommunications, in 2018 at 122.0-122.25 GHz and 244-246 GHz \cite{CEPT18a}. 

\begin{figure}    
	\centering
	\includegraphics[width=0.5\textwidth]{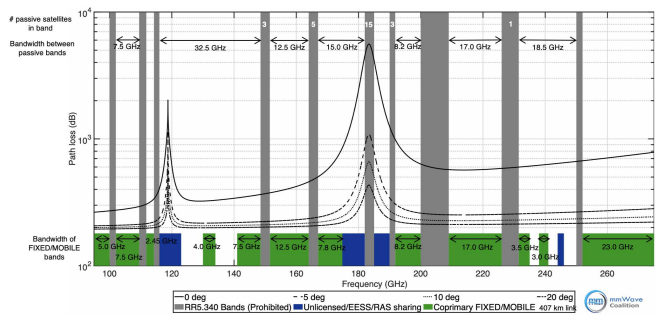}
	\caption{Key ITU spectrum allocation above 100 GHz with RR5.340 prohibited bands, unlicensed/EESS/RAS sharing bands, and coprimary fixed/mobile bands, from Millimeter Wave Coalition. }
	\label{fig:mmWc}
				\vspace{-2.0 em}
\end{figure}

In March 2019, the Federal Communications Commission (FCC) adopted the Notice of Proposed Rulemaking ET Docket 18-21 \cite{FCC19a} and four new unlicensed bands were authorized (116-122, 174.8-182, 185-190, and 244-246 GHz), shown as the blue segments along the bottom of Fig. \ref{fig:mmWc}. The Office of Communications, the communication regulator of the UK, published a statement document in October 2020 on supporting innovation in the 100-200 GHz bands \cite{ofcom20}, and opened over 18 GHz of radio spectrum across three bands similar to the FCC (116-122, 174.8-182, and 185-190 GHz). 

\textcolor{black}{As seen in Fig. \ref{fig:mmWc}, the International Telecommunications Union (ITU) Radio regulation 5.340 (RR5.340 which was adopted at the ITU World Radio Conference in 2000, WRC-2000) prohibits any emission in ten passive bands (the gray bars) to protect satellite sensors and deep space observatories from 100 GHz up to 252 GHz  \cite{marcus21a}.} The black double arrows show the bandwidth between RR5.340 bands, where the largest chunk of contiguous spectrum blocks available is 32.5 GHz (between 116-148.5 GHz) out of the total available 180 GHz (between 95-275 GHz) \cite{rappaport19access,FCC19a}. Other shared bands have less severe restrictions, such as the green segments in the bottom of Fig. \ref{fig:mmWc} where Fixed and Mobile services have coprimary allocations. The ITU agreed in Resolution 731 to study if and under what conditions sharing the ten forbidden bands with terrestrial networks is possible. 


\textcolor{black}{This letter explores the fundamental free space and rain-related radio propagation characteristics of the sub-THz and THz frequency bands (see \cite{xing19GC,rappaport19access,xing21icc,ma18channel,abbasi20ICC,nguyen2018comparing} that cover other propagation issues such as scattering, multipath or path loss at 140 GHz). This letter also demonstrates how the engineering efforts to create mmWave 5G networks will carry forward to frequencies as high as 800-900 GHz, meaning that the engineering developments of adaptive beamforming, wideband channel allocation, and site-specific installation used to create today's 5G networks will hold for the coming several decades as mobile communications move up into the THz range.} Recent results of rooftop surrogate satellite/tower base station measurements in downtown Brooklyn, NY, at 142 GHz show the isolation between terrestrial networks and surrogate satellite systems (as well as between mobiles and the base station for integrated access and backhaul). This isolation will stem from designing mobile transmitters and antennas to always maintain their main radiation energy lobe close to the horizon. Results here motivate the exploration of innovative ways to enable new terrestrial products and services to occupy the vast spectrum above 100 GHz in a manner that protects the incumbent passive satellites and space-based sensors. This work could lead to effective spectrum utilization and breakthroughs in coexistence techniques, without interfering with incumbent satellite sensors, or at least providing confidence that out-of-band emission (OOBE) levels in adjacent bands would not be harmful to passive satellite sensors above 100 GHz.

\section{Atmospheric and rain attenuation at frequencies above 100 GHz}~\label{sec:Atmo}
\begin{figure}    
	\centering
	\includegraphics[width=0.45\textwidth]{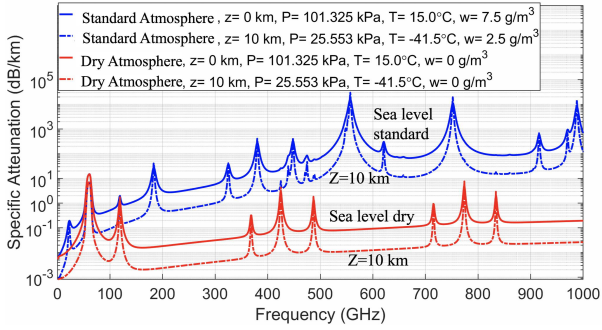}
	\caption{Atmospheric absorption beyond Friis free space path loss \cite{friis1946note} at z = 0 km height (sea level) and z= 10 km height (10 km above the sea level), computed from models in \cite{ITU-Rattenuation}.}
	\label{fig:AtmoAtt}
				\vspace{-1.5 em}
\end{figure}

Atmospheric absorption of sub-THz and THz transmissions is greater than frequencies below 6 GHz (where air causes attenuation of only fractions of dB/km), but the difference is not as great as most believe \cite{rappaport19access,ITU-Rattenuation}. Certain frequency bands such as 183, 325, 380, 450, 550, and 760 GHz (see Fig. \ref{fig:AtmoAtt}) suffer much greater attenuation beyond the free space propagation loss over distances due to atmospheric absorption \cite{rappaport19access,ITU-Rattenuation}, which makes these bands well suited for very short range and secure communications on earth (e.g., whisper radios or future WiFi above 100 GHz) as well as key frequencies for space sensors and telescopes operating above the earth's atmosphere to detect the presence of atomic elements \cite{802.15.3d,rappaport19access,ma18channel}. 

The atmospheric attenuation is highly related to the altitude above earth, air pressure, temperature, and water vapor density. Fig. \ref{fig:AtmoAtt} shows the atmospheric absorption beyond the natural Friis free space path loss (FSPL) from \cite{friis1946note} at sea level (z = 0 km) and for the channel operating at 10 km above the sea level (z= 10 km) in both dry conditions (e.g., desert where water vapor density $w$ is close to 0 $g/m^3$) and standard conditions (e.g., $w=7.5~g/m^3$). The air pressure, temperature, and water vapor density decrease when the altitude increases (the relationship can be found in \cite{ITU-Rattenuation}), which results in larger atmospheric absorption at sea level than higher up in the troposphere compared to today's 4G networks below 6 GHz (e.g., but only about 6 dB/km at 300 GHz, z=0 km) and even less attenuation at higher altitudes (e.g., only about 1 dB/km at 300 GHz, z=10 km). This dispels myths, and shows air attenuation is inconsequential up to THz.

\begin{figure}    
	\centering
	\includegraphics[width=0.38\textwidth]{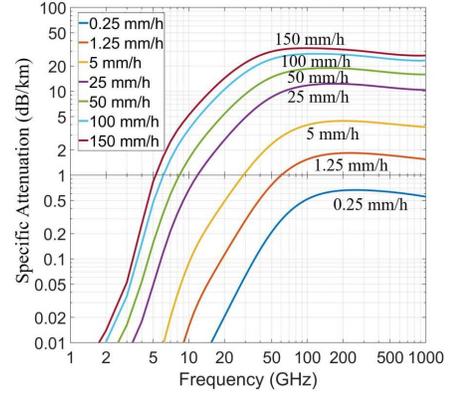}
	\caption{Rain attenuation beyond natural FSPL in dB/km over frequency at various rainfall rates using ITU models \cite{ITU-Rattenuation,qingling2006rain,rappaport2013millimeter}.}
	\label{fig:rainATT}
			\vspace{-1.0 em}
\end{figure}

\textcolor{black}{Notably, the atmospheric absorption at sea level in standard condition (the blue curve in Fig. \ref{fig:AtmoAtt}) at 200-300 GHz is remarkably less than 10 dB/km, and even at 800-900 GHz the additional atmospheric absorption beyond the natural Friis free space loss is 100 dB/km at sea level, meaning only 10 dB per 100 m over today's 4G cellular, which will be compensated for by the antenna gains at higher frequencies \cite{rappaport19access}.} Our companion paper in this issue \cite{xing21b} shows that within office buildings, there is remarkable similarity in terms of large-scale path loss exponents when going from 28 GHz to 140 GHz, when referenced to the first meter of free-space propagation \cite{Sun16b,Samimi15b}, implying that the THz channels are very similar to today's mmWave wireless propagation channels except for the path loss in the first meter of propagation when energy spreads into the far field \cite{rappaport19access,Sun16b,xing21b,Samimi15b}. Fig. \ref{fig:rainATT} shows that there is a vast amount of spectrum up to 1 THz with relatively little attenuation that can be utilized for future mobile and fixed terrestrial communication.

\begin{figure}    
	\centering
	\includegraphics[width=0.45\textwidth]{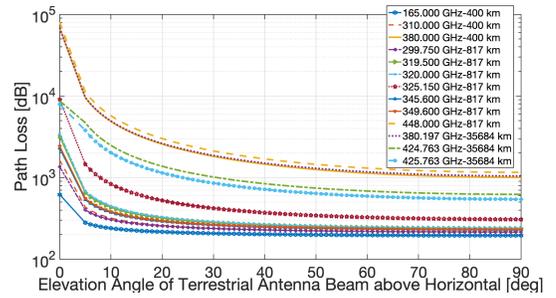}
	\caption{Total path loss from ground terminals to satellite NGSO EESS altitudes (without antenna gains) computed from models in \cite{ITU-Rattenuation,coalition19b}.}
	\label{fig:PLSate}
		\vspace{-1.5 em}
\end{figure}

\begin{figure}    
	\centering
	\includegraphics[width=0.45\textwidth]{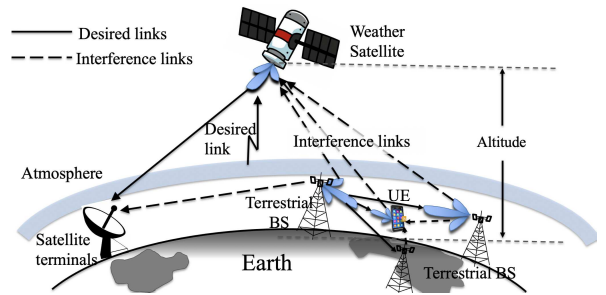}
	\caption{Satellite and Terrestrial networks, illustrating the interference between weather satellites (passive) and terrestrial communication (active) systems.}
	\label{fig:RT1}
	\vspace{-1.5 em}
\end{figure}

\textcolor{black}{Fig. \ref{fig:PLSate} shows how the earth's troposphere offers a natural impenetrable wall to interference in space at low elevation angles (large slant paths). Atmospheric attenuation rapidly increases to hundreds and even thousands of dB when low angles are used from earth to orbiting satellites (see Fig. \ref{fig:PLSate}) at various satellite altitudes and frequency bands from 165-425 GHz.} This natural attenuation provided by the earth's atmosphere (mainly from the troposphere) to orbiting satellites above 100 GHz is remarkably effective provided that earth emissions are kept low on the horizon (15\textdegree~or less) as Fig. \ref{fig:mmWc} and Fig. \ref{fig:PLSate} demonstrate.

\begin{figure}    
	\centering
	\includegraphics[width=0.4\textwidth]{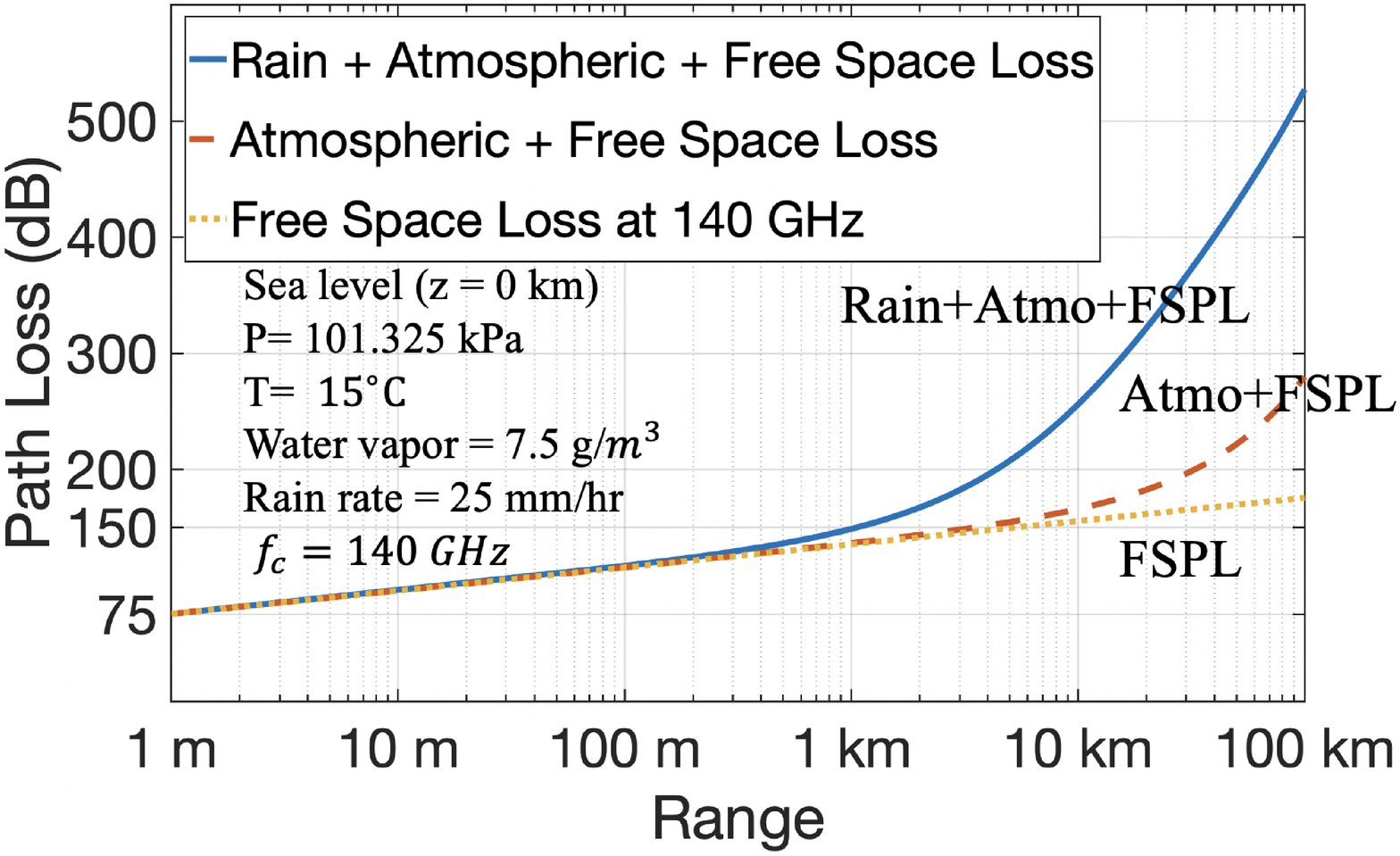}
	\caption{Total path loss at 140 GHz (0\textdegree~elevation angle) without antenna gains including free space path loss (FSPL), rain attenuation, and atmospheric absorption on a terrestrial path using models in \cite{ITU-Rattenuation}.}
	\label{fig:TotPL}
		\vspace{-1.5 em}
\end{figure}

The rain attenuation beyond the natural Friis free space loss \cite{rappaport19access,friis1946note} across frequency at various rainfall rates is shown in Fig. \ref{fig:rainATT}, indicating that above 70 GHz, further increases in frequencies are \textit{not} further impacted by rain. This is encouraging since coverage distances in today's 5G mmWave networks will not be hampered by rain as carrier frequencies move up to THz \cite{rappaport2013millimeter,rappaport19access}. It is worth noting that the ITU-R rain attenuation model \cite{ITU-Rspecific} is used to compute data shown in Fig. \ref{fig:rainATT}, however, the Crane Model \cite{crane96rain} is used in \cite{rappaport19access}, which predicts a 15 dB/km greater loss at extremely heavy rain rates near 150 mm/hr above 100 GHz. Work in \cite{hirata09a} shows the ITU-R model fits well with measurement data at rainfall rate up to 80 mm/h, however, more measurement data are needed at rainfall rates near 150 mm/h. 

Total path loss including FSPL, rain attenuation, and atmospheric absorption for fixed backhaul (energy beamed close to the horizon) at 140 GHz (at sea level with the standard conditions) is shown in Fig. \ref{fig:TotPL}, indicating the atmospheric absorption has remarkably little impact on total path loss out to about 10-20 km, although heavy rain will practically limit fixed THz links to several km. 

\section{Rooftop Surrogate Satellite Measurements and Interference Analysis}
\textcolor{black}{Results in Section \ref{sec:Atmo} are very encouraging and illustrate the feasibility of future terrestrial networks in sub-THz and THz bands. However, there is still a dearth of knowledge about frequencies above 100 GHz, and extensive radio propagation measurements and realistic channel models are needed to support future system design.} Currently, most research above 100 GHz has focused on very close range measurements (e.g., less than 10 m) or only line-of-sight (LOS) scenarios due to the difficulty of achieving sufficient transmit power and measurable ranges \cite{Ju20a}. Early work on urban and indoor channel characteristics are found in \cite{Ju20a, xing19GC,xing21icc,ma18channel,abbasi20ICC,nguyen2018comparing}.

To our knowledge, the world's first air-to-ground measurements at 142 GHz were conducted in Downtown Brooklyn from September 2020 to December 2020, which can be used for emulating satellite-to-ground and UAS/HAPS-to-ground communications. This measurement campaign was designed to give early insights for spectrum sharing/coexistence techniques and interference between terrestrial networks and surrogate satellite systems (or for terrestrial networks that will use mobiles and tall base stations for integrated access and backhaul). This measurement campaign used the channel sounder system described in \cite{xing19GC}. Fig. \ref{fig:Mea2Loc} shows the rooftop base station (BS) receiver and ground user equipment (UE) transmitter locations on NYU's downtown Brooklyn campus.

\subsection{Rooftop Surrogate Satellite Measurements at 142 GHz}
\begin{figure}    
	\centering
	\includegraphics[width=0.45\textwidth]{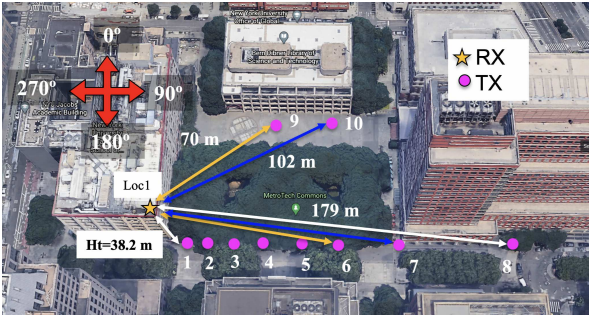}
	\caption{Rooftop surrogate satellite measurement campaign. The surrogate satellite (and backhaul) receiver RX location is at 38.2 m above the ground on the rooftop identified as a yellow star. Ten mobile TX locations on the ground are identified as purple circles. The LOS elevation pointing angles from TX1-8 to the RX location are 80\textdegree~to 15\textdegree, respectively.}
	\label{fig:Mea2Loc}
			\vspace{-1.0 em}
\end{figure}

In the rooftop surrogate satellite measurements, the BS (RX) was placed on the rooftop of a nine-story building which is 38.2 m above ground, emulating a passive satellite receiver. Horn antennas with 8\textdegree~half power beam width (HPBW) were used at both the rooftop RX and ground-based TX, where the antennas were mechanically steered and extensively rotated to consider all possible pointing combinations in the search of energy. In satellite communications, the received interference power level will be highly dependent on the ground-based transmitter's radiation pattern and any multipath that is reflected or scattered up to the satellite. The satellite will view wide swaths of earth, such that any radiation source from earth would add to others for a cumulative interference effect in the satellite's antenna pattern which may be mitigated by the massive attenuation in the troposphere shown in Fig. \ref{fig:PLSate}. To study the variation of received power with elevation angles in a realistic urban setting, eight ground-based TX (1.5 m above the ground) locations were chosen in the NYU Brooklyn courtyard (see Fig. \ref{fig:Mea2Loc}), having LOS elevation angles in 10\textdegree~decrements ranging from 80\textdegree~to 10\textdegree. Due to the space limitation of the measurement area, the farthest TX provided a 15\textdegree~elevation angle boresight to the RX instead of 10\textdegree. The channel sounder requires a clear LOS link for calibration, but TX locations 1 through 8 were somewhat blocked by tree foliage. To overcome this issue, two additional TX locations - TX 9 and 10 were chosen, which had the same link lengths as TX 6 and 7, respectively, for free space calibration without any link obstructions \cite{xing21icc}.

While foliage would further attenuate energy from mobiles on earth received at a satellite or backhaul system, it could also serve as a source of scattering, so the experiments were designed to carefully try to detect any energy whatsoever in any possible boresight direction used on the ground and roof-based RX.

A multipath power delay profile (PDP) for the (sometimes foliage-blocked) LOS boresight TX-RX pointing combination was first measured at each TX location, and then the TX was rotated 360\textdegree~in the azimuth plane by steps of 8\textdegree~(the HPBW of the antennas used at both the TX and RX), and this was repeated at elevation angles of 0\textdegree, 8\textdegree, 16\textdegree, 24\textdegree, and 32\textdegree. For each TX pointing angle, an exhaustive manual search was conducted at the roof-mounted RX to attempt to capture any signals (e.g., direct path, reflected, or scattered rays).  

\subsection{Measurement Results}~\label{sec:results}
\begin{figure}    
	\centering
	\includegraphics[width=0.5\textwidth]{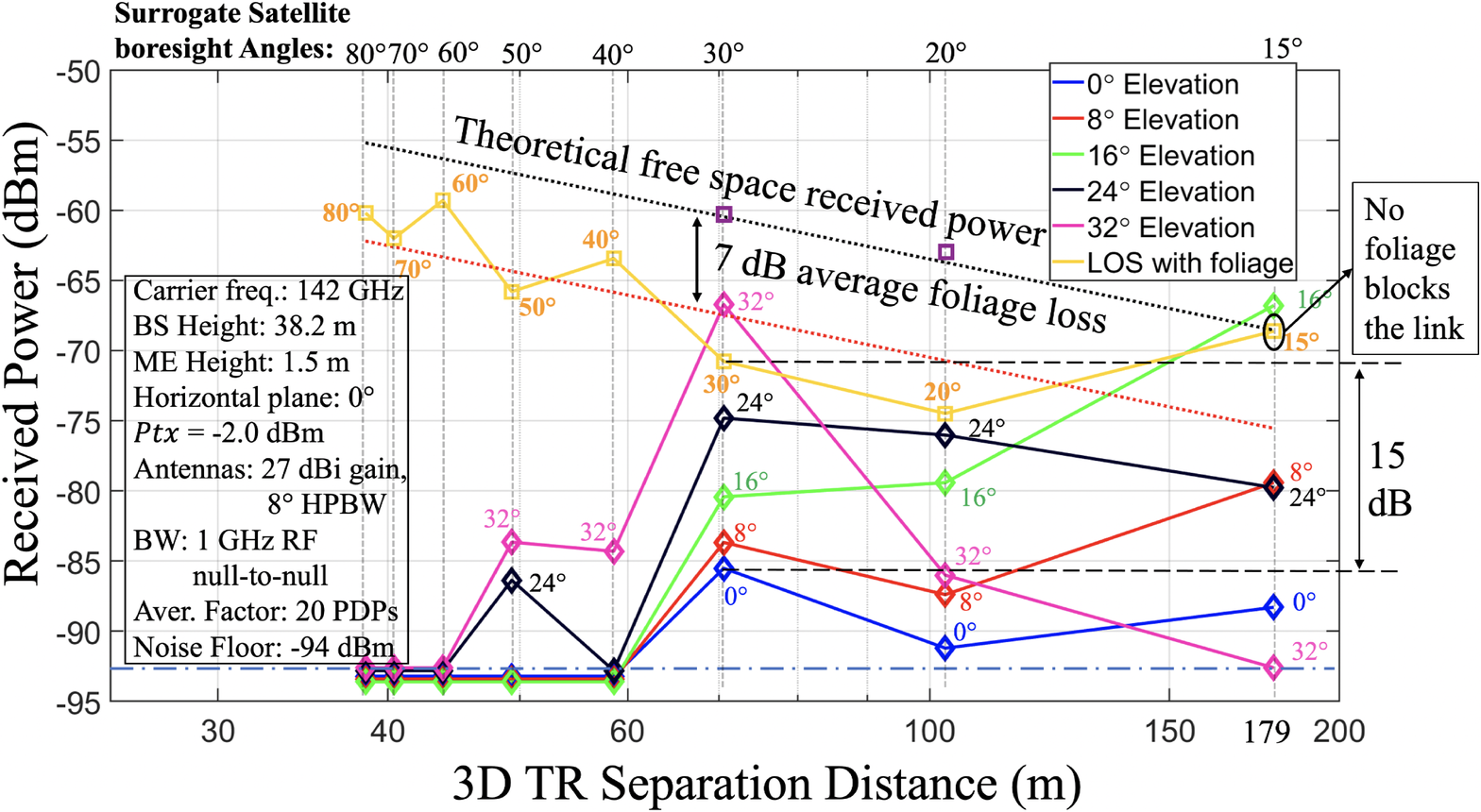}
	\caption{The rooftop base station (38.2 m above the ground) received power vs. different distances and different elevation angles form ground users (1.5 m above the ground) at 142 GHz. }
	\label{fig:RTPR}
			\vspace{-2.0 em}
\end{figure}

\textcolor{black}{Fig. \ref{fig:RTPR} presents the measured receiver power at the rooftop RX (38.2 m above the ground) from the ground-based TXs (1.5 m above the ground) at different elevation angles and distances from 40 to 180 m, corresponding to TX1-8 as shown in Fig. \ref{fig:Mea2Loc}. The 0\textdegree~elevation angle signifies the horizontal plane and the positive values represent the TX elevation angles above the horizon.} The yellow curve shows the best-fit measured power of the foliage-blocked LOS links between the ground TX and roof mounted RX, and reveals 7.1 dB foliage loss beyond free space at 142 GHz with a standard deviation of 3.7 dB about the best fit since the foliage loss is affected by the wind.

When the ground-based TX antenna is pointing at a 0\textdegree~elevation angle on the horizon (the blue curve in Fig.\ref{fig:RTPR}), there is virtually no power captured by the rooftop RX even when the ground TX has antenna pattern energy leaking from its antenna pattern while pointing nearly directly to the roof (the boresight elevation angle is 15\textdegree). The worst case of interference was found when the TX is at Location 6 (70 m) and Fig.\ref{fig:RTPR} shows how raising the elevation angle of the ground based transmitter dramatically increases energy detected by the roof-mounted RX, due to antenna pattern leakage and multipath from surrounding buildings.

\subsection{Analysis of Interference between Satellite and Terrestrial Networks}

Obviously, this measurement campaign and Fig. \ref{fig:RTPR} is cursory and does not consider the massive attenuation due to the troposphere slant path. As shown in Figs. \ref{fig:mmWc} and \ref{fig:PLSate}, the slant paths close to the horizon (e.g., 0-15\textdegree) experience much more atmospheric absorption and path loss due to longer path length within the troposphere than with overhead paths (e.g., 90\textdegree~elevation angle). Thus, if antenna patterns of earth transmitters are carefully designed (e.g., adaptive antenna patterns with very low sidelobes overhead), the passive receiver in the air (satellite) will not receive interference from the ground, enabling spectrum sharing between satellites and ground terminals. This approach would limit OOBE, as well.

Analysis of the interference from ground transmitters to the TEMPEST-D satellite receiver at 165 GHz, one of the major satellite systems working at sub-THz frequency bands \cite{TEMPEST16b} was further considered to determine preliminary satellite interference levels from a mobile system on earth. The required sensitivities of the TEMPEST-D satellite radiometric passive sensors are astounding, only $\Delta T_e = $ 0.1 Kelvin, implying 7 dB SNR (permissible interference level at 20\% of the required sensitivity $\Delta P [\text{W}]= k \Delta T_e B $ \cite{ITU-RS.2017}) would occur for a interference level of -133.0 dBm over $B=$ 200 MHz, $I \text{[dBm]} = 10\log_{10} (k \Delta T_e B \times 10^3) - \text{SNR} = -133.0 $ dBm, where $k$ is the Boltzmann's constant $= 1.38 \times 10^{-23} \text{J/K}$, and $I$ is the noise/interference level (see Table II in \cite{ITU-RS.2017}).

The signal between a ground TX and a satellite undergoes several stages of propagation and attenuation \cite{3GPP38.811}:
\begin{equation}
	\small
\label{equ:NTNPL}
PL= PL_b + PL_g + PL_s,
\end{equation}
where $PL$ is the total path loss in dB, $PL_b$ is the basic propagation path loss in dB, $PL_g$ is the atmospheric gasses attenuation in dB, and $PL_s$ is the attenuation due to either ionospheric or tropospheric scintillation in dB. The basic propagation path loss is modeled as:

\begin{equation}
	\small
\label{equ:FSPL}
PL_b = \text{FSPL} (h/\sin(\alpha),f_c) + SF + CL (\alpha,f_c),
\end{equation}
where $\text{FSPL}(h/\sin(\alpha),f_c)$ is the free space path loss in dB, $SF$ is a log-normal distributed shadow fading in dB, $CL (\alpha,f_c)$ is the cluster loss in dB (e.g., foliage loss, building penetration loss), $h/\sin(\alpha)$ is approximately the link distance in meters, $\alpha$ is the elevation pointing angle in degrees from the ground terminals to the satellite, and $f_c$ is the carrier frequency in Hz \cite{3GPP38.811}. When the TX is in LOS of the satellite, the cluster loss is negligible in \eqref{equ:FSPL}.

Assuming the ground-based TX is in LOS of the satellite and transmits its main beam at an elevation angle of $\alpha=10$\textdegree, the basic propagation loss $PL_b$ in \eqref{equ:NTNPL} at $f_c=165$ GHz with $h=400$ km altitude is 204.0 dB. The atmospheric gasses attenuation $P_g$ in \eqref{equ:NTNPL} is $\sim35.2$ dB \cite{ITU-Rattenuation}. Assuming reasonable parameters for a mobile ground station transmitting a 200 mW signal at 165 GHz \cite{rodwell21a} using a 15 dBi gain antenna \cite{simsek20a}, the theoretical received power at the TEMPEST-D radiometer would be -201.2 dBm which is more than 60 dB (a factor of 1,000,000) below the minimum signal detection level at the satellite (-133 dBm over 200 MHz). This implies that if $N$ mobile devices were operating on the ground, each with similar transmitter power levels and all $N$ devices were within the passband and main beam of the satellite receiver, the total contribution of the entire terrestrial network interference power $I$ (NOTE: variable capital $I$ for interference power, assuming each interference adds power non-coherently) is $I \text{[dBm]}  = -201.2 \text{dBm} + 10 \log_{10} (N) =$ -133.0 dBm. Solving for $N$ yields $N = 6.6$ million mobile ground devices would sum up to be equal to the noise floor seen by the satellite receiver, therefore not causing deleterious effects. If the satellite was overhead or not near the horizon, properly designed mobile antennas could provide tens of dB additional attenuation that could offset the smaller tropospheric loss on overhead paths.

Improved antenna patterns (e.g., spatial filtering) at both the ground and the satellite need to be considered, and foliage and building blockage (for indoor systems) would provide greater protection than this simple example shows. Note that we did not consider any rain/fog/cloud attenuation, foliage/building penetration loss, the impact of in-building use or any antenna pointing offset issue which would further attenuate the ground-based transmissions. The analysis indicates that with proper transmit power limits and antenna designs, active terrestrial mobile communications could possibly not interfere with the passive satellite applications at frequencies above 100 GHz, although much more study is needed to ensure protection of the expensive and ultra-sensitive satellites after they are launched.

\textcolor{black}{Note that the interference analysis in this letter is focused on frequencies above 100 GHz between the active terrestrial networks and satellite passive receivers in space, and the downlink of satellite networks which is generally below 40 GHz is not considered.} This analysis gives some insights into the key issues for spectrum sharing and suggest that current regulations that prohibit any ground-based transmitters in bands above 100 GHz may be too restrictive in light of the intense attenuation of sub-THz and THz frequencies through the troposphere over long slant-paths. The spectrum mask (out-of-band emission limits for ground-based transmitters) can be properly designed as a function of the elevation angle, based on this analysis for the passive satellite frequency bands above 100 GHz \cite{coalition19b}. 

\section{Conclusion}
This paper presents recent global spectrum regulations as well as fundamental atmospheric and rain attenuation considerations at frequencies above 100 GHz which show there is no fundamental physical channel impediment for utilizing sub-THz and THz bands up to 1 THz for future wireless communications. Rooftop surrogate satellite measurements at 140 GHz are presented, showing the isolation between the terrestrial networks and surrogate satellite systems as well as the isolation between terrestrial mobile users and co-channel fixed backhaul links. The surrogate satellite measurements and preliminary analysis suggest that propagation on the horizon (e.g., elevation angles $\leq$ 15\textdegree) may not cause interference (same or adjacent bands) between passive satellite sensors and terrestrial transmitters at frequencies above 100 GHz if the antenna patterns of the transmitters are carefully designed to avoid radiation in space (e.g., adaptive antenna patterns with very low sidelobes). This work offers a critical first step in addressing the spectrum co-existence challenges across the entire mmWave and THz RF ecosystem to spur technology development in future terrestrial networks in the sub-THz and THz bands.

\bibliographystyle{IEEEtran}
\bibliography{Indoor140GHz}

\end{document}